\documentclass[14pt]{article}

\usepackage{arxiv}
\usepackage{caption}
\usepackage[utf8]{inputenc} % allow utf-8 input
\usepackage[T1]{fontenc}    % use 8-bit T1 fonts
\usepackage{hyperref}       % hyperlinks
\usepackage{url}            % simple URL typesetting
\usepackage{booktabs}       % professional-quality tables
\usepackage{amsfonts}       % blackboard math symbols
\usepackage{nicefrac}       % compact symbols for 1/2, etc.
\usepackage{microtype}      % microtypography
\usepackage{graphicx}
\usepackage{amsmath}
\graphicspath{ {./images/} }

\title{AI-Driven Accelerated Discovery of Intercalation-type Cathode Materials for Magnesium Batteries}

\author{
  Wenjie Chen \\
  Department of Physics \\
  Tsinghua University\\
  Beijing, 100086, China\\
  \texttt{chenwj22@mails.tsinghua.edu.cn} \\
   \And
  Zichang Lin \\
  Department of Physics \\
  Tsinghua University\\
  Beijing, 100086, China\\
  \texttt{lzc23@mails.tsinghua.edu.cn} \\
  \And
  Xinxin Zhang\\
  Department of Physics \\
  Tsinghua University\\
  Beijing, 100086, China\\
  \texttt{zhangxx19@mails.tsinghua.edu.cn} \\
  \And
  Hao Zhou\\
  Institute of AI Industry Research \\
  Tsinghua University\\
  Beijing, 100086, China\\
  \texttt{zhouhao@air.tsinghua.edu.cn} \\
  \And
  Yuegang Zhang\\
  Department of Physics \\
  Tsinghua University\\
  Beijing, 100086, China\\
  \texttt{yuegang.zhang@tsinghua.edu.cn} \\
}

\begin{document}
\maketitle
\begin{abstract}
Magnesium-ion batteries hold promise as future energy storage solution, yet current Mg cathodes are challenged by low voltage and specific capacity. Herein, we present an AI-driven workflow for discovering high-performance Mg cathode materials. Utilizing the common characteristics of various ionic intercalation-type electrodes, we design and train a Crystal Graph Convolutional Neural Network model that can accurately predicts electrode voltages for various ions with mean absolute errors (MAE) between 0.25 and 0.33 V. By deploying the trained model to stable Mg compounds from Materials Project and GNoME AI dataset, we identify 160 high voltage structures out of 15,308 candidates with voltages above 3.0 V and volumetric capacity over 800 Ah/L. We further train a precise NequIP model to facilitate accurate and rapid simulations of Mg ionic conductivity. From the 160 high voltage structures, the machine learning molecular dynamics simulations have selected 23 cathode materials with both high energy density and high ionic conductivity. This AI-driven workflow dramatically boosts the efficiency and precision of material discovery for multivalent ion batteries, paving the way for advanced Mg battery development. 
\end{abstract}

% keywords can be removed

\section{Introduction}
Magnesium-ion batteries (MIBs), using an earth-abundant, safe Mg anode of high volumetric capacity (3833 $\mathit{mAh}\mathit{cm}^{-3}$, versus 2046 $\mathit{mAh}\mathit{cm}^{-3}$ of Li anode), have received widespread attention\cite{levi_phase_2006,yoo_mg_2013}. However, searching advanced magnesium cathode materials with improved voltage, energy density, and fast ion transport capability is urgently needed. In recent years, the rapid expansion of databases based on density functional theory (DFT) has enabled the application of machine learning in the screening of electrode materials. The emergence of innovative algorithms, such as SchNet\cite{schutt_schnet_2018},Materials Graph Network (MEGNet)\cite{chen_graph_2019}, and Crystal Graph Convolutional Neural Network(CGCNN)\cite{xie_crystal_2018}, have significantly boosted the precision and efficiency of material screening. Thesemethods have achieved great success in lithium-ion batteries (LIBs), or instance, for prediction of cathode voltage\cite{zhang_interpretable_2022}, ionic conductivity of solid-state electrolytes\cite{sendek_holistic_2017}, and battery state-of-charge in electrical vehicles\cite{hong_multi-_2024}. However, existing battery datasets for deployment of these models are predominantly composed of LIB data (Supplementary Fig. 1), hindering the screening of electrode materials for other working ions.

\begin{figure}
  \centering
  \includegraphics[width=0.95\textwidth]{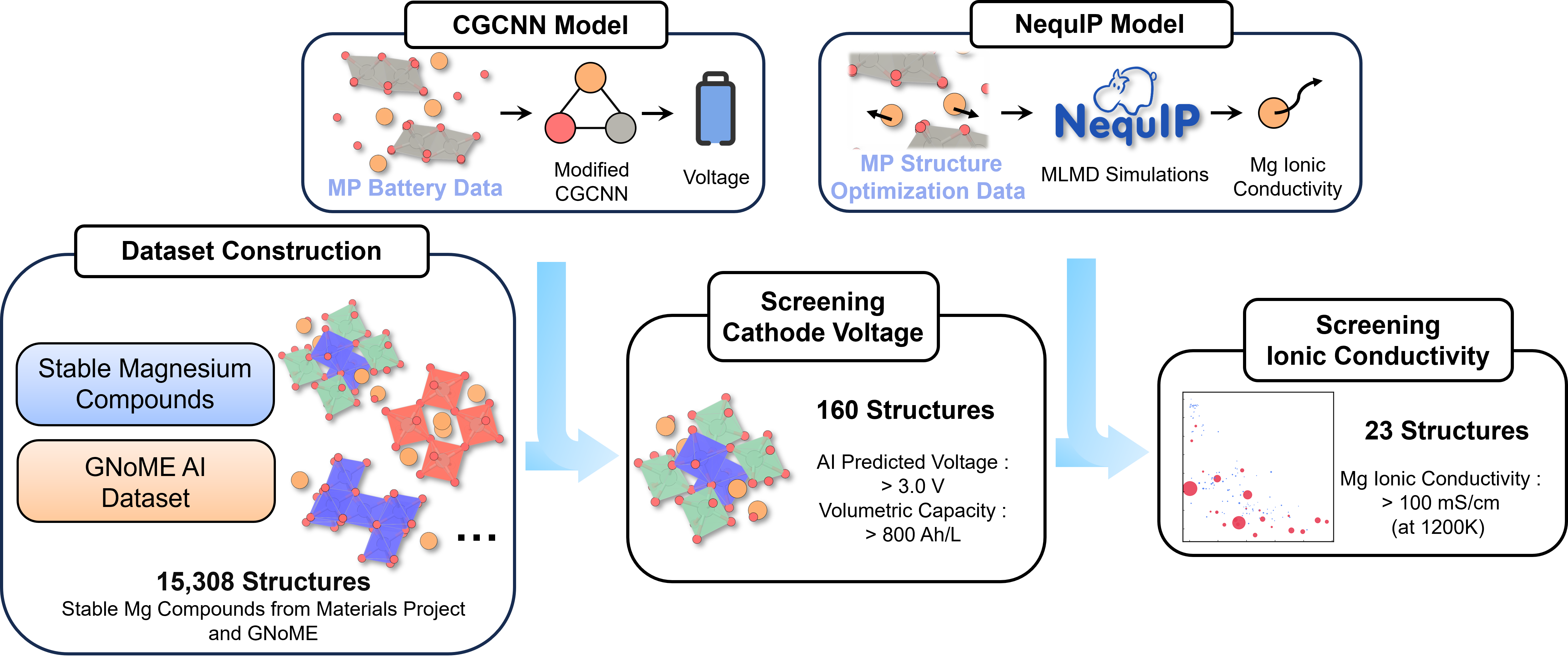}
  \caption{\textbf{Illustration of the end-to-end AI workflow.} We utilize battery data from the Materials Project to train a CGCNN model specifically designed for predicting cathode voltage. This model is then applied to a dataset of 15,308 magnesium compound candidates from MP and GNoME, selecting 160 high-voltage structures. Then, we train a NequIP model using the structure optimization data from MP and apply it to the AI-predicted high-voltage structures. Finally, 23 materials with both high voltage and high ionic conductivity emerge as promising Mg cathode candidates.}
  \label{fig:fig1}
\end{figure}

In this work, we develop an end-to-end AI-driven discovery workflow for electrode materials of Mg batteries, as shown in Fig. \ref{fig:fig1}. First, we train a universal CGCNN model that is capable of predicting the voltage of all intercalation-type electrodes. Unlike previous reports that use transfer learning to overcome the challenge of small datasets\cite{zhang_interpretable_2022}, our model utilizes the common characteristics of the intercalation-type electrode materials, enabling us to train all electrodes of different working ions simultaneously. By this way, we significantly reduce the mean average error (MAE) of the voltage for Li, Mg, Na, Al, K, Ca, and Zn
cathodes to 0.32 V, 0.29 V, 0.32 V, 0.33 V, 0.30 V, 0.29 V and 0.25 V, respectively. Then, we deploy this model on stable Mg compounds from Materials Project (MP) dataset and Graph Networks for Materials Exploration (GNoME) AI database, selecting Mg cathode candidates with voltage above 3.0 V and volumetric capacity over 800 Ah/L\cite{jain_commentary_2013,merchant_scaling_2023}. First-principles calculations are then introduced to validate the prediction. For the selected high-voltage materials, we further employ Neural Equivariant Interatomic Potentials (NequIP) to perform machine learning molecular dynamics (MLMD) simulations, and calculate their Mg ionic conductivity\cite{batzner_e3-equivariant_2022}. This AI-driven workflow overcomes the limitations of small datasets, enabling us to pinpoint 23 promising candidates from a pool of 15,308 options for new intercalation-type magnesium cathode materials with high energy density and ionic conductivity. This paves the way for further experimental study and industrial development of Mg batteries.

\section{Result and Discussion}
\label{sec:headings}
\subsection{Screening the Voltage of Mg Cathode Materials}

\begin{figure}
  \centering
  \includegraphics[width=0.75\textwidth]{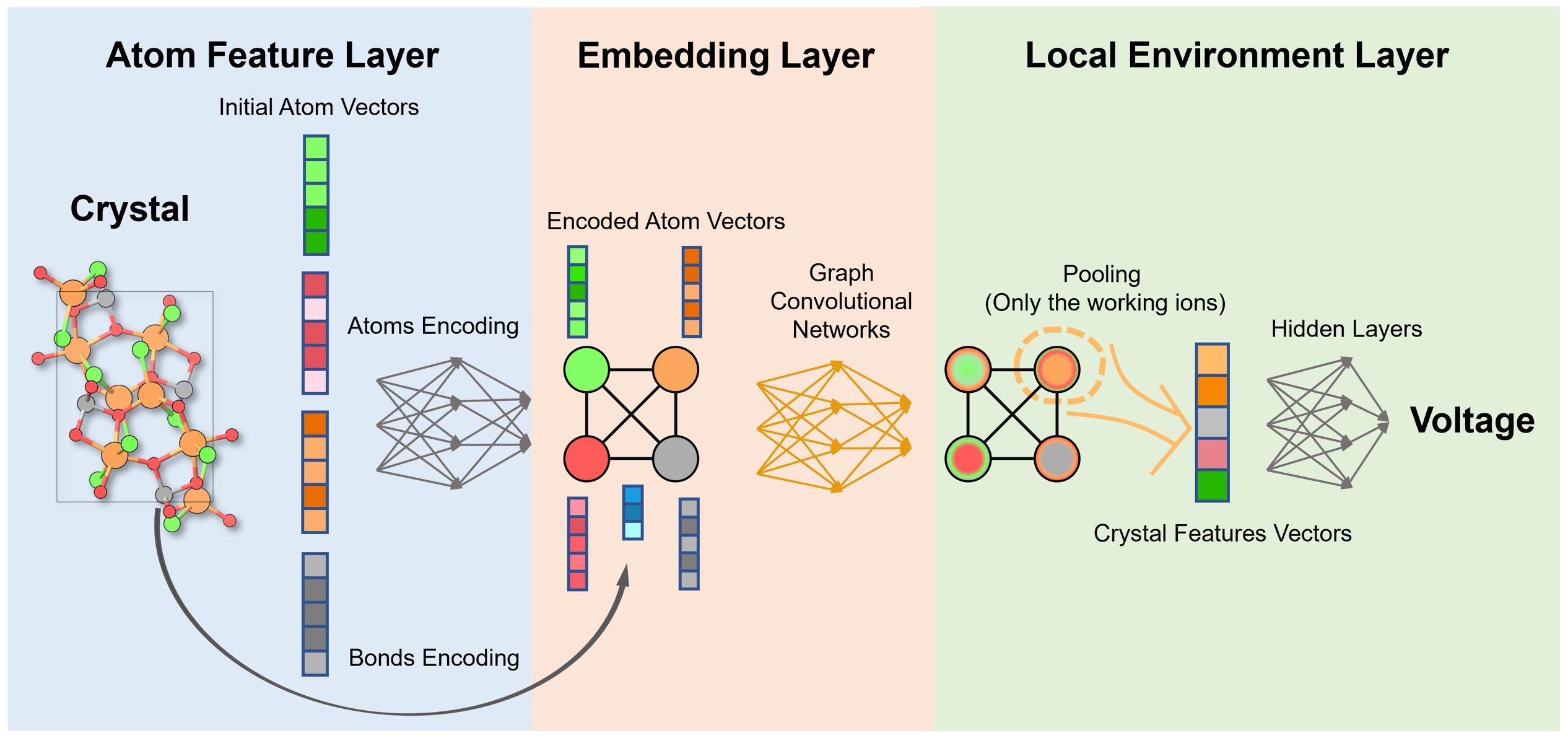}
  \caption{\textbf{Illustration of the modified CGCNN model for voltage prediction.} A crystal is represented as a multigraph G where atoms are embedded into the nodes and the connection between atoms are embedded into edges. Only working ions are pooled and used to predict the electrode voltage.}
  \label{fig:fig2}
\end{figure}

We introduce CGCNN model and modify it to match the prediction of electrode voltage. The model of CGCNN in our work is illustrated as Fig. \ref{fig:fig2}. In this model, a crystal is represented as a multigraph where atoms are embedded into the nodes and the connection between atoms are embedded into edges. Then, three layers of graph convolution are introduced to update the atom feature vectors based on its surrounding environment. Unlike the typical CGCNN that pools all nodes to represent the entire crystal, our modified network specifically pools the vectors of the working ions, followed by an output layer that predicts the cathode voltage. This approach allows us to utilize datasets from all intercalation-type cathodes as input, since the open cell voltage of a cathode material is determined by the change in free energy during deintercalation:

\begin{equation}
E=-\frac{\mathit{\Delta G}}{\mathit{nF}}\\
\end{equation}

where E refers to electrode voltage,  $\mathit{\Delta G}$ denotes the change of Gibbs free energy when an ion is transferred from cathode to metal anode, n is the charge of the transferred ion, and F is the Faraday constant. It's noteworthy that the change in the Gibbs free energy mainly depends on the local environment of the working ion, which is well represented through the process of graph convolution in the model that aggregate the surrounding information into the atom vector of the working ion. This approach in Fig. 2 captures the common characteristics of various ionic intercalation-type cathodes and enables us to use fully discharged electrode structures to construct our dataset. This provides a more comprehensive understanding compared to previous studies that dealt with cathodes for different working ions separately. By applying graph convolution, the working ions acquire detailed information about their electrochemical environment, which is essential for accurate voltage prediction.

\begin{figure}
  \centering
  \includegraphics[width=0.95\textwidth]{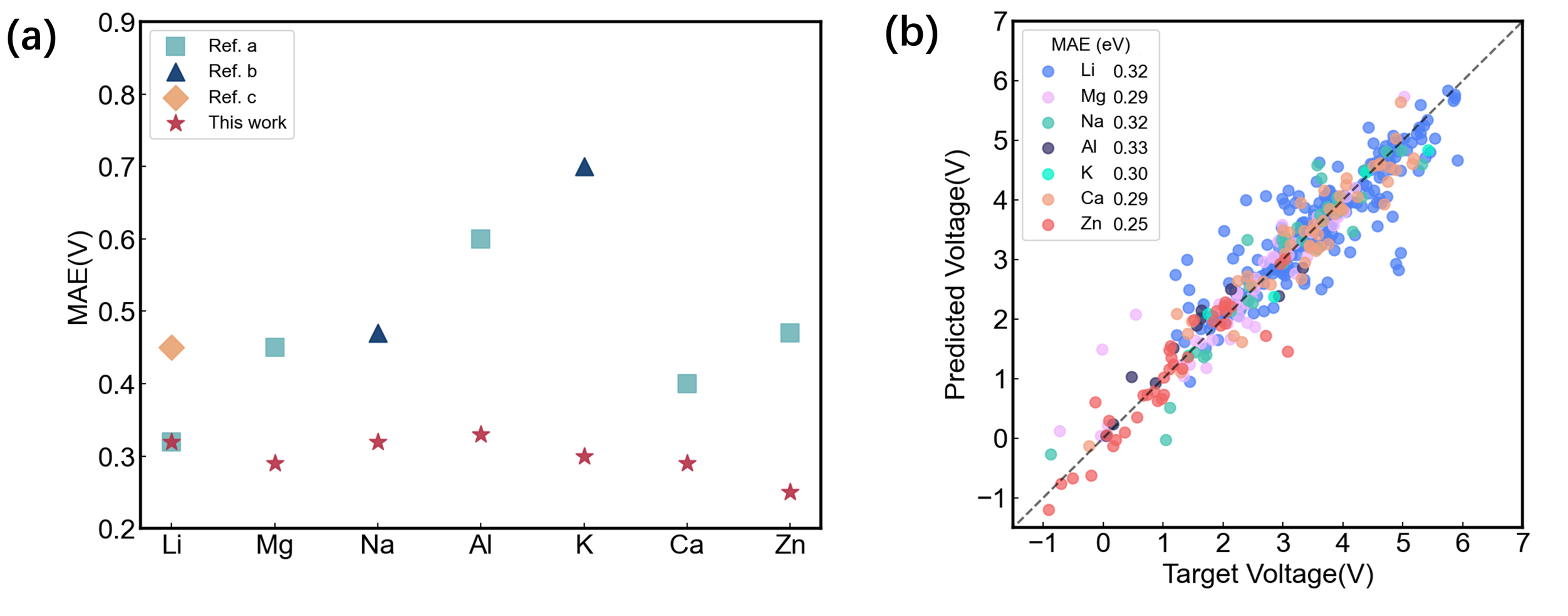}
  \caption{\textbf{ Machine Learning performance. (a)}  Comparison of the MAE for all working ions with previous studies. Ref. a\cite{zhang_interpretable_2022} Ref. b\cite{louis_accurate_2022} Ref. c\cite{dinic_unconstrained_2023}. \textbf{(b)} Plots of predicted voltage and target voltage in test set.}
  \label{fig:fig3}
\end{figure}

A total of 4,045 samples of different working ions from the Battery Explorer of Materials Project are used for the training of our modified CGCNN\cite{zhou_first-principles_2004}. Samples are randomly split into training, validation, and test sets with the ratio of 0.8:0.1:0.1, while maintaining the same proportion for each working ion. Mean Absolute Error (MAE) is employed as the loss function and the evaluation metric. After fully training, the MAE of our models for Li, Mg, Na, Al, K, Ca, and Zn cathodes are 0.32 V, 0.29 V, 0.32 V, 0.33 V, 0.30 V, 0.29 V and 0.25 V, respectively, which outperform previous reports on predictions for LIBs and multivalent metal-ion batteries, as depicted in Fig. \ref{fig:fig3}(a)\cite{zhang_interpretable_2022,dinic_unconstrained_2023,louis_accurate_2022}. The model shows a consistent prediction MAE for all ionic electrodes, independent of their respective dataset size, highlighting its ability to effectively harness the similarities inherent in intercalation-type materials. The predicted voltages plotted against target voltages of the test set are distributed around dashed line y=x, as shown in Fig. \ref{fig:fig3}(b) and Supplementary Fig. 2, indicating a high accuracy of our model for all types of batteries. 

\begin{figure}
  \centering
  \includegraphics[width=1\textwidth]{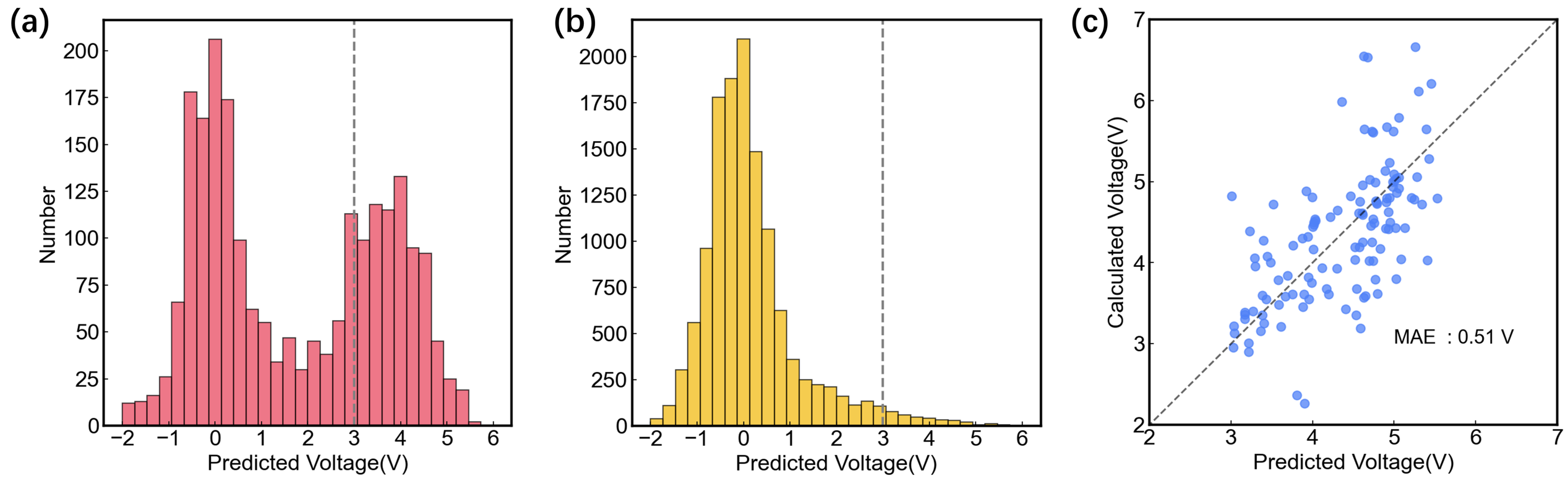}
  \caption{\textbf{ Voltage screening and first-principles validation for high-voltage Mg cathode materials.} Predicted voltage distribution for compounds from \textbf{(a)} Materials Project and \textbf{(b)} GNoME database. \textbf{(c)} First-principles validation for selected materials with predicted voltage higher than 3V (Supplementary Table 2).}
  \label{fig:fig4}
\end{figure}

To further verify the model's capability, we applied it to a combination of two datasets: stable Mg compounds sourced from the Materials Project (2,202 structures with an energy above hull lower than 150 meV/atom) and the GNoME AI database (13,106 structures with negative formation energy, see Supplementary Table 1). It is evident that the introduction of the GNoME AI dataset has significantly broadened the scope of material screening. Voltage distribution of these two datasets is shown in Fig. \ref{fig:fig4}(a) and \ref{fig:fig4}(b), respectively. We find it intriguing that the MP database exhibited a bimodal voltage distribution around 0 V and 4 V, whereas the predicted voltages for GNoME database cluster around 0V. This difference may originate from the different criteria for choosing stable structures in the two databases. A total of 160 structures with predicted voltages higher than 3.0 V and volumetric capacities above 800 Ah/L are selected, and the first-principles calculations were performed to verify their predicted voltages. The comparison of predicted voltage against first-principles calculated voltage is illustrated in Fig. \ref{fig:fig4}(c), showing the high accuracy (MAE of 0.51V) of our model for voltage prediction. 

\begin{figure}
  \centering
  \includegraphics[width=0.95\textwidth]{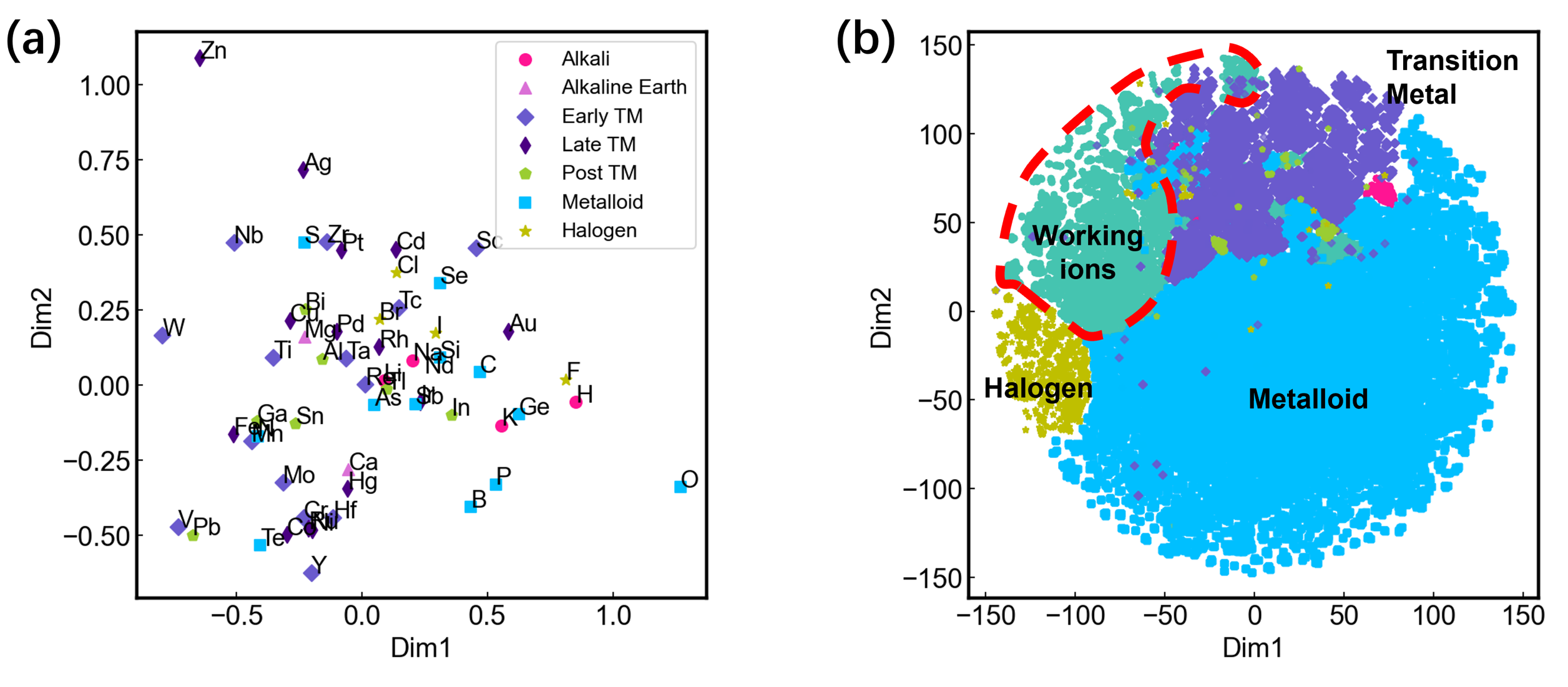}
  \caption{\textbf{Hierarchical visualization of the CGCNN model. (a)} The PCA-reduced 2-dimensional space of the atomic feature layer for each element. \textbf{(b)} The t-SNE distribution of the atoms in the local environment layer where each class of the elements are grouped together. The working ions (Li, Mg, Al, Na, K, Ca, Zn) are circled with a dashed red line.}
  \label{fig:fig5}
\end{figure}

The impressive performance of our CGCNN model in predicting voltages of various electrode materials stems from its powerful capability to extract and comprehend atomic-level information\cite{xie_hierarchical_2018}. A general feature of CGCNN model is its capability for hierarchical visualization, which enables us to visualize the compositional and structural similarities between materials in an arbitrary material space with representations learned from different layers of the networks\cite{xie_hierarchical_2018}. To further demonstrate this feature, here, we adjust the initial atom feature vectors into random binary vectors of length 640 to ensure no element information is passed to the model (Supplementary Fig. 3). In the embedding layers, the atom features learned by our model are set to vectors of length 64, where principal component analysis (PCA) is introduced to project these vectors to 2-dimensional space\cite{hotelling_relations_1992}. As depicted in Fig. \ref{fig:fig5}(a), all elements are randomly scattered on x-y plane, indicating that the positional information of elements in the periodic table has not been learned due to random input of atomic features. However, after applying three layers of GCN to aggregate local chemical environments, the t-SNE \cite{van_der_maaten_visualizing_2008} dimensionality-reduced data shows these atoms in specific
chemical environments are effectively grouped together according to their element classes, such as halogen, metalloid and transition metal (TM) elements, as shown in Fig. \ref{fig:fig5}(b). Additionally, working ions (Li, Mg, Al, etc.) from various types of batteries cluster together in the local environment layer. This indicates that the model can extract atomic
features from random encodings during the training process, making the effective training of various ionic types of electrodes possible. 

\subsection{Screening the ionic conductivity of High-voltage Mg Cathode Materials}

Ionic conductivity is a key parameter for cathode materials as it significantly influences the overall efficiency and performance of a battery\cite{fu_materials_2023,cui_cathode_2024}. However, screening ionic conductivity of cathode materials using high precision ab-initio MD method can be challenging due to the computational capability and cost limitations. MD simulation with machine learning force field is a way to address this problem without sacrificing too much precision. Herein, we utilize NequIP model and the structural relaxation data from Materials Project to construct robust, precise and universal force fields for Mg cathode materials, and employ it to screen ionic conductivity for high-voltage cathodes materials selected in the previous section\cite{batzner_e3-equivariant_2022}. To enable fast and stable training of NequIP model, only the first three and the final steps of the structural relaxations for 151,336 crystals in the Materials Project are utilized. This sampling method is proved to be effective, because most intermediate steps are similar to each other and involve small forces, not useful for efficient training\cite{merchant_scaling_2023}. After a full training, our model can accurately predict the energy and force with MAE of 40.8 meV/atom and 61.8 meV/Å, respectively, which is comparable to existing models (Supplementary Fig. 4)\cite{merchant_scaling_2023,xie_gptff_2024}.

To calculate ionic conductivity of the selected high-voltage Mg cathode materials, we create a supercell for each structure that initially contains at least 512 atoms, but with one-fifth of the Mg atoms deleted randomly to create ionic migration interstices. We then perform 10ps of NPT ensemble NequIP-based MLMD simulations at 1200K to equilibrate lattice vectors, followed by 50 ps of NVT ensemble simulations at 1200K to achieve Mg ionic conductivity. All structures successfully pass this simulation sequence without any crash, demonstrating the excellent stability of our model. The diffusivity of Mg ions is calculated through:

\begin{equation}
D=\frac 1{2\mathit{d\Delta t}}\frac 1 N\sum _i^N\left[r_i\left(t+\mathit{\Delta
t}\right)-r_i\left(t\right)\right]^2\\
\end{equation}

where d=3 represents the dimension of the supercell,  $\mathit{\Delta t}$ is the time interval and  $r_i(t)$ represent the position of ion $i$ out of N ions at time $t$. We can then obtain the ionic conductivity $\sigma $ through Einstein Relation:

\begin{equation}
\sigma =\frac{Ne^2Z^2}{Vk_BT}D\\
\end{equation}

where N is the total number of ions,  $Z=2$ for magnesium ions, V is volume and T is temperature.

\begin{figure}
  \centering
  \includegraphics[width=0.95\textwidth]{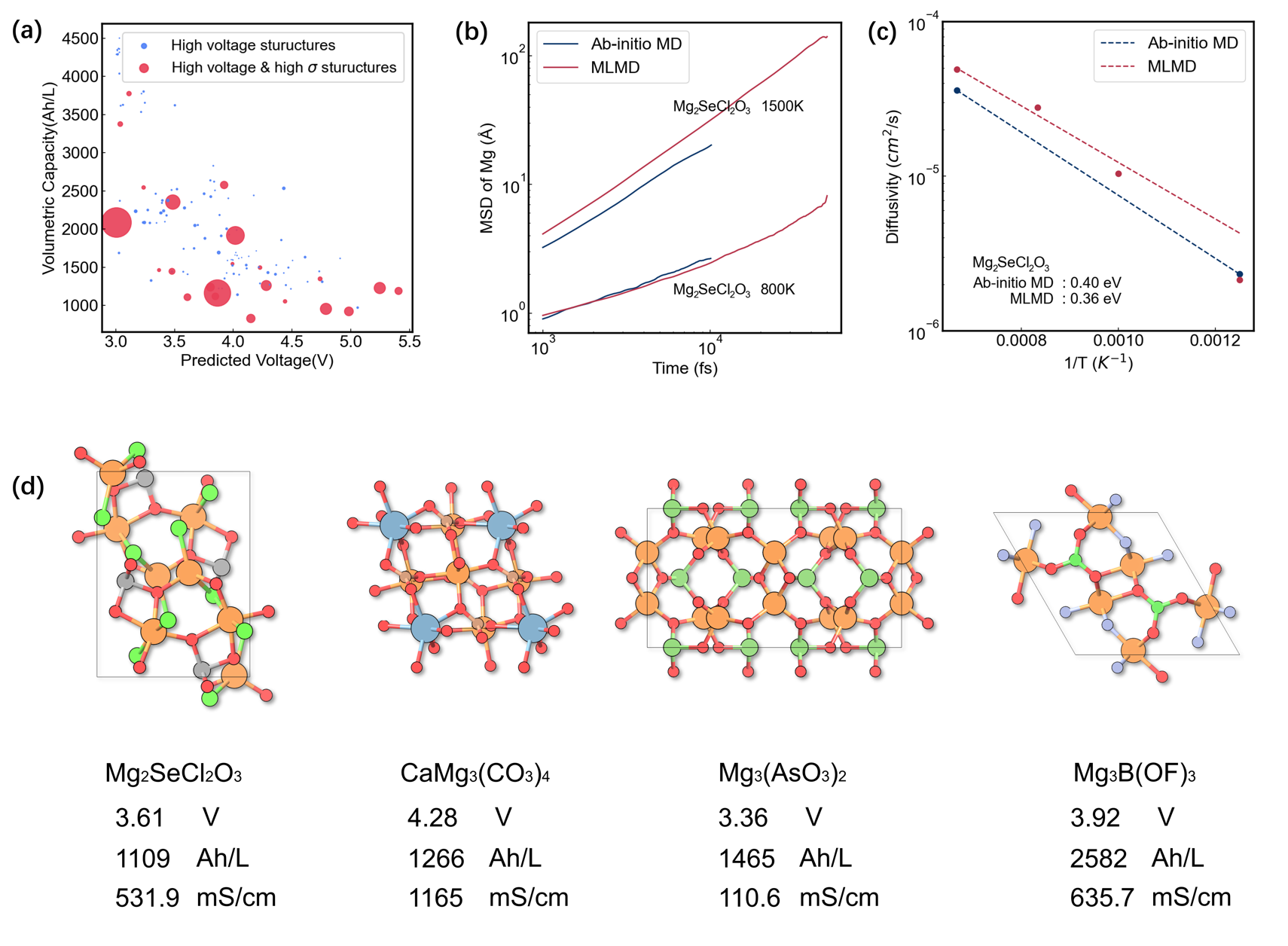}
  \caption{\textbf{MLMD screening of Mg ionic conductivity. (a)} The plot of the selected high-voltage Mg cathode materials, where the size of the data point corresponds to the relative ionic conductivity calculated using NequIP-based MLMD. The materials with an ionic conductivity greater than 100 $\mathit{mS}{\cdot}cm^{-1}$ are highlighted in red. \textbf{(b)} Comparison of Mg MSD results calculated by MLMD and ab-initio MD for $\mathit{Mg}_2\mathit{SeCl}_2O_3$ at 800K and 1500K. \textbf{(c)} Arrhenius plot of Mg diffusivity as a function of temperature for ab-initio MD and MLMD. \textbf{(d)} Examples of selected candidates with high voltage, volumetric capacity and Mg ionic conductivity.}
  \label{fig:fig6}
\end{figure}

The MLMD results for the selected high-voltage Mg cathode materials at 1200K are shown in Fig. \ref{fig:fig6}(a), where the x-axis shows the CGCNN-predicted cathode voltage, y-axis represents the calculated volumetric capacities, and the size of the scattered points corresponds to the relative ionic conductivity calculated using MLMD. The materials with an ionic conductivity greater than 100 $\mathit{mS}{\cdot}cm^{-1}$ are screened out and highlighted in red in Fig. \ref{fig:fig6}(a). To validate the accuracy of this MLMD model, we select a high-voltage, high-mobility material, $\mathit{Mg}_2\mathit{SeCl}_2O_3$, for ab-initio MD validations. Considering the high computational cost of ab-initio MD calculations, we construct a supercell of $4\times 2\times 2$ and randomly delete one-fifth of the Mg atoms, and perform NVT ab-initio MD at 800K and 1500K for 10 ps. For MLMD, we use the same supercell created in the previous section and perform 50 ps NVT simulations at 800K, 1000K, 1200K and 1500K, respectively. The comparison of mean square displacement (MSD) of Mg ions between the MLMD and ab-initio MD trajectories is shown in Fig. \ref{fig:fig6}(b), indicating very similar ion migration behaviors using these simulation methods. Fig. \ref{fig:fig6}(c) presents the Arrhenius plots of Mg diffusion coefficients at various temperatures simulated by MLMD and ab-initio MD. The diffusion barriers derived from ab-initio MD (0.40 eV) and MLMD (0.36 eV) are remarkably close, demonstrating the NequIP-based MLMD model's high accuracy. So far,we have screened thousands of Mg structures and identified 23 materials with a voltage higher than 3.0V, a volumetric capacity greater than 800Ah/L, and ionic conductivity over 100 $\mathit{mS}{\cdot}cm^{-1}$ at 1200K. Examples of these structures are shown in Fig. \ref{fig:fig6}(d) and Supplementary Table 3. These selected high-voltage Mg cathode structures hold great promise in future research of high-energy-density Mg batteries.

\section{Conclusion}

In summary, we have developed an end-to-end AI-driven workflow designed to explore materials with high cathode voltage and high ionic conductivity. This workflow excels predicting the cathode voltage across various working ions, reaching an accuracy of 0.29 V for magnesium electrodes. We have identified 160 high voltage structures with voltages above 3.0V and volumetric capacity over 800Ah/L out of 15,308 candidates. By training a NequIP model to facilitate accurate and rapid MLMD simulations for Mg ionic conductivity screening, we have further identified 23 Mg cathode materials that have both high energy density and high ionic conductivity. By integrating future experiment validation, we hope our method for screening magnesium electrode materials could pave the way for accelerating development of Mg batteries.

\paragraph{Computational Details}
\bigskip
\paragraph{Training of Crystal Graph Convolutional Neural Networks.} We use modified Crystal Graph Convolutional Neural Networks (CGCNN) to predict cathode voltage directly from crystal structures\cite{xie_crystal_2018}. In CGCNN, a crystal is represented as a multigraph  $G$ where atoms are embedded into nodes and the connections between atoms within a cutoff radius are embedded into edges. The model learns the features of the multigraph by passing messages between nodes through graph convolutional layers. In this work, we use default initial atom vectors  $v_i^{(0)}$ from CGCNN for training (Supplementary Fig. 3). Distances between atoms are encoded into edge feature vectors of length 41:

\begin{equation}
u_{\left(i,j\right)_k}\left[t\right]=\exp \left(\frac{-\left(d_{\left(i,j\right)_k}-\mu
_t\right)^2}{\sigma ^2}\right)\\
\end{equation}
where $d_{\left(i,j\right)k}$ denotes the distance between i and j at their $k^{\mathit{th}}$  edge, and $\mu _t=t\cdot 0.2$ Å for t = 0, 1, . . ., 40. The combined atom feature vectors 
 $z_{\left(i,j\right)_k}^{\left(t\right)}=v_i^{\left(t\right)}{\oplus}v_j^{\left(t\right)}{\oplus}u_{\left(i,j\right)_k}$ are then processed through three convolution layers:

\begin{equation}
v_i^{\left(t+1\right)}=v_i^{\left(t\right)}+\sum _{j,k}\sigma
\left(z_{\left(i,j\right)k}^{\left(t\right)}W_f^{\left(t\right)}+b_f^{\left(t\right)}\right){\odot}g\left(z_{\left(i,j\right)k}^tW_s^{\left(t\right)}+b_s^{\left(t\right)}\right)\\
\end{equation}

where ${\odot}$ denotes element-wise multiplication, $\sigma $ stands for sigmoid function, and $g$ is a non-linear activation function. W and b refer to weights and biases of the neural networks, respectively. After three layers of convolution, an average pooling function is applied specifically to the working ions to generate an overall feature vector $v_c$ for the cathode material. We can finally map the crystal structure to a voltage value
through a hidden layer after pooling.

\paragraph{First-principles Validation of High Voltage Cathode Materials.} We utilize Density Functional Theory (DFT) to calculate voltages of Mg cathode materials\cite{PhysRev.140.A1133}. All density functional calculations are performed using the Vienna Ab Initio Simulation Package (VASP)\cite{PhysRevB.54.11169}. The potentials are constructed using the projector augmented wave (PAW) pseudopotential within the Perdew-Burke-Ernzerhof (PBE) exchange-correlation functional framework\cite{perdew_generalized_1996,PhysRevB.59.1758}. A cutoff energy of 520 eV and a resolution of the Monkhorst-Pack k-point mesh of 0.2 $\text{Å}^{-1}$ are used in electronic structure calculations\cite{PhysRevB.13.5188}. The convergence criteria for electronic relaxation is set to $10^{-6}\mathit{eV}$ between two self-consistent steps. For cell relaxation, the criterion is 0.01 eV/Å for all atoms. A Hubbard U extension is added to the GGA Hamiltonian (GGA+U) using the reference value on Materials Project to eliminate the spurious self-interaction errors in the d-electrons\cite{zhou_first-principles_2004,jain_formation_2011}. The voltage of the intercalation-type cathode can be calculated by equation (1). Although computing Gibbs free energy with DFT calculations is time-consuming, the difference between energy and Gibbs free energy is approximately 25 meV for crystals at roomtemperature\cite{louis_accurate_2022}. This enables us to use the following equation to calculate the voltages of Mg cathodes:

\begin{equation}
V_{\mathit{Mg}}=\frac{E_{\text{charged}}-E_{\text{discharged}}-n{\cdot}E_{\text{Mg}}}{2n{\cdot}F}\\
\end{equation}

where $E_{\text{charged}}$($E_{\text{discharged}}$) is the total energy of the cathode material in the charged (discharged) state. $E_{\text{Mg}}$ is the energy of a Mg atom in bulk magnesium, and n represents the number of de-intercalated magnesium atoms. Given that our calculations focus on intercalation-type electrode materials, we use DFT-D3 for energy corrections to better estimate the deintercalation energy\cite{10.1063/1.3382344}. VASPKIT is employed for batch data preprocessing\cite{wang_vaspkit_2021}.

\paragraph{Training of Neural Equivariant Interatomic Potentials}We utilize Neural Equivariant Interatomic Potentials (NequIP) for batch molecular dynamics simulation of high voltage cathode materials\cite{batzner_e3-equivariant_2022}. We build a four-layer model with hidden features of 128 $\ell $ = 0 scalars, 64 $\ell $ = 1 vectors and 32 $\ell $ = 2 tensors (or irreducible representations $128\times 0e+64\times 1e+32\times 2e$ for short), and edge-irreducible representations of 0e + 1e + 2e. A radial cutoff of 5 Å is used for interatomic distances embedding. To achieve fast and effective training, we employ a dataset comprising 567,815 frames, which includes only the first three steps and the final step of the structural optimization trajectories for all crystals from the Materials Project. Models are trained with the Adam optimizer using a learning rate of $2\times 10^{-3}$ and a batch size of 32. Huber loss functions are employed with the weights of 1:5 for energy
and force. After 38 epochs of training, the MAE for energy and force are 40.8 meV/atom and 61.8 meV/Å, respectively.

\section*{\textbf{Acknowledgments}}

This work was supported by the National Key R\&D Program of China (2022YFA1203400) and the National Natural Science Foundation of China (W2441009). The authors thank Dr. Xu Yong for helpful discussion.

\bibliographystyle{unsrt}

\bibliography{references}

\end{document}